\documentclass[apj]{emulateapj}

\usepackage{color}

\definecolor{darkgreen}{rgb}{0.0,0.6,0.0}

\newcommand{\be}{\begin{equation}}
\newcommand{\ee}{\end{equation}}
\newcommand{\bea}{\begin{eqnarray}}
\newcommand{\eea}{\end{eqnarray}}
\def\se#1{Section~\ref{sec:#1}}

\def\Fig#1{Figure~\ref{#1}}
\def\Table#1{Table~\ref{#1}}
\def\Eq#1{Eq.~(\ref{#1})}

\begin{document}

\title{Revised Mass-to-Light Ratios for Nearby Galaxy Groups and Clusters}

\author{Yutong Shan$^{1,2,*}$, Michael McDonald$^{3,4}$, and St\'ephane Courteau$^1$}
\altaffiltext{1}{Department of Physics, Engineering Physics and Astronomy, Queen's University, Kingston, ON, Canada}
\altaffiltext{2}{Now at Harvard-Smithsonian Centre for Astrophysics, 60 Garden Street, Cambridge, MA 02138}
\altaffiltext{3}{Kavli Institute for Astrophysics and Space Research, MIT, Cambridge, MA 02139, USA}
\altaffiltext{4}{Hubble Fellow}
\altaffiltext{*}{Email: yshan@cfa.harvard.edu}

\begin{abstract}
We present a detailed investigation of the cluster stellar mass-to-light ($M^*/L$) ratio and cumulative stellar masses, derived on a galaxy-by-galaxy basis, for 12 massive ($M_{500} \sim 10^{14} - 10^{15} M_\sun$), nearby clusters with available optical imaging data from the Sloan Digital Sky Survey Data Release 10 and X-ray data from the {\it{Chandra X-ray Observatory}}.  Our method involves a statistical cluster membership using both photometric and spectroscopic redshifts when available to maximize completeness whilst minimizing contamination effects. We show that different methods of estimating the stellar mass-to-light ratio from observed photometry result in systematic discrepancies in the total stellar masses and average
mass-to-light ratios of cluster galaxies. Nonetheless, all conversion methodologies point to a lack of correlation between $M^*/L_i$ and total cluster mass, even though low-mass groups contain relatively more blue galaxies.  We also find no statistically significant correlation between $M^*/L_i$ and the fraction of blue galaxies ($g-i < 0.85$). For the mass range covered by our sample, the assumption of a Chabrier IMF yields an integrated $M^*/L_i \simeq 1.7 \pm 0.2$ M$_\sun$/L$_{i,\sun}$, a lower value than used in most similar studies, though consistent with the study of low-mass galaxy groups by \citet{lea12}. A light (diet) Salpeter IMF would imply a $\sim$60\% increase in $M^*/L_i$. 
\end{abstract}


\section{Introduction}\label{sec:intro}
Given their large size ($\sim$Mpc), mass ($\sim10^{14}$ M$_{\odot}$), and energetics ($>10^{64}$ ergs in cluster-cluster mergers), galaxy clusters make excellent laboratories for a variety of extragalactic and cosmological studies \citep[see, e.g.,][for a review]{voit05}.  Of particular interest to galaxy formation modelling is trying to understand the relative distribution and evolution of the three main contributors to cluster mass: the stars (primarily residing in galaxies), the hot intracluster medium (ICM), and the dark matter.  The dominant dark matter and gas components are fairly well constrained by gravitational lensing \citep[e.g.,][]{kaiser95,allen98,bartelmann01,high12,vonderlinden12} and X-ray observations \citep[][]{white97,jones99,mohr99,allen02,allen04,vik06}, respectively. However, the stellar mass, as traced by optical or near-IR emission, is often poorly determined, with typical uncertainties exceeding factors of two \citep[see e.g.,][]{gon07,gon13,courteau14}.  Uncertain stellar mass-to-light ($M^*/L$) ratios, which reflect the nature of stellar populations in various galaxy environments and their dependencies on global cluster parameters, are the dominant source of error in stellar mass estimates. Additional uncertainty comes from the large extent of the central cluster galaxy which blends into the faint and diffuse intracluster light \citep[e.g.,][]{gonz05, kra14}. 

Precise mass accounting of the baryons and dark matter can provide constraints on cosmological parameters, under the assumption that the average cluster's composition should be representative of the cosmic proportions \citep[e.g.,][]{whi93,allen02,voit05}. Thus, one expects the baryon fraction, $f_{baryon}=(M^*+M_{gas})/M_{tot}$, as well as the stellar mass fraction, $f_\star=M^*/M_{tot}$, to be common to clusters of all scales \citep{sch06}, and within $\sim$10\% of the universal mean (e.g. \citealt{kra05}). However, current estimates of the baryon fraction in galaxy clusters fall short of the universal mean $f_b = \Omega_b / \Omega_m \approx 0.15$ and 0.17 measured by Planck \citep{Planck13} and WMAP \citep{dun09, kom11} respectively, with multiple studies showing that the stellar fraction, $f_\star$, decreases with increasing cluster mass \citep[e.g.,][]{lin03,gon07,gio09,and10,dai10,lag11,zha11,lin12,lea12,gon13}. Moreover, these authors find that the star-to-gas fraction is a strongly decreasing function of cluster mass \citep[see Fig.~10 in][]{gon13}. 
The reliable interpretation of these trends relies critically on the accuracy and applicability of the adopted mass measures.  Multiple examples of the so-called ``uniform-field approximation'' \citep{whi93} whereby a single {\emph{constant}} $M^*/L$ ratio is used to compute the total stellar masses of galaxy clusters are found throughout the literature. In addition to failing to explicitly consider the variation in stellar and galactic populations within a cluster, the adopted $M^*/L$ values are typically derived from studies of biased galaxy populations (e.g. often applicable to an old ($\sim10$Gyr) stellar population). 

A standard approach to retrieving galaxy properties (including stellar masses) from spectra uses stellar population synthesis (SPS) modelling \citep{tin-gun76,tin78,bru83,bc93,mar05,conroy13,
courteau14}.  
This approach involves synthesizing theoretical stellar spectra and fitting a library of ensemble scenarios to the observed galaxy spectra or spectral energy distribution (SED) in order to assign parameters of the best-fit population model to the physical galaxy (i.e. SED-fit). The standard ingredients of SPS are stellar evolution models (e.g. \citealt{bc03, mar05}) and an initial mass function \citep[IMF; e.g.,][]{salpeter55, kro01,cha03}. The evolution of multiple, single stellar populations (SSPs) are superimposed over one another according to a prescribed set of stellar formation histories (SFHs; see \citet{wal11}, \citet{conroy13}, and \citet{courteau14} for recent reviews). The accuracy of the stellar population model typically improves as the spectral coverage increases, but so do the observational and computational expenses.

The relationship between colour and $\log(M^*/L)$ (or ``CMLR'' for `Colour Mass-to-Light Relations') has been determined for various widely-used broadband filter combinations (e.g. \citealt{beldJ01,bel03,zib09,tay11,ip13}).  While robust within their respective definitions, the scatter between these CMLRs is still large due to intrinsic sample variance, model degeneracies, observational limitations, and more. Systematic differences between CMLRs arise from implementing different stellar evolution models using a range of assumptions for, e.g., the IMF, SFH, and metallicity, as well as different SPS libraries, many of which are contentious \citep{courteau14}. Even with the highest quality multi-wavelength data, these various uncertainties cap the accuracy of CMLR stellar mass determinations at $\sim 0.1 - 0.3$ dex \citep{gal-bel09, con09, con-gun10, beh10}. Other possible avenues to rapidly estimating $M^*/L$, such as relations between $M^*/L$ and galaxy luminosity (or ``LMLR'' for `Luminosity Mass-to-Light Relations'), carry comparable uncertainties \citep[e.g.][]{vdM91, bel03, kauf03,cap06}. Despite their limitations, these methods have gained popularity for their straightforward implementation over large galaxy samples.

For instance, the \citet{cap06} LMLR is the basis of the stellar mass estimates made by \citet{gon07} and \citet{and10}.  In \citet{gon07}, the variation in cluster galaxy population is taken to be encoded in the luminosity function (LF), which is itself uncertain. Integration over the cluster LF results in a luminosity-weighted $M^*/L_I = 3.6$. \citet{and10} use an average value from \cite{cap06} of $M^*/L_I = 3.8$. However, as \citet{lea12} point out, the SAURON sample upon which the \citet{cap06} LMLR is based comprises only early-type galaxies. A dynamical mass component, which includes contributions from dark matter, is also invoked in modeling their LMLR . \citet{gon13} correct for the latter effect but still adopt a single, constant (albeit revised and more representative) mass-to-light ratio for all clusters ($M^*/L_I = 2.65$). \citet{lin03} employ a similar method, utilizing separate $M^*/L_K$'s for ellipticals and spirals weighted by their respective K-band luminosity functions. The final cluster $M^*/L_K$ is an average of the two galaxy populations, weighted by the expected spiral fraction for that X-ray temperature ($T_X$).   Other examples of recent stellar mass census that rely on mass-to-light ratios include \citet{arn07,dai10,zha11,lin12}.
\citet{lea12} caution against the use of constant $M/L$ ratios in converting cluster luminosity into stellar masses, arguing that these are biased towards certain galaxy populations. Using a sample of galaxy groups in the COSMOS multi-waveband survey ($10^{13}M_\sun < M_{halo} < 10^{14}M_\sun$), they showed that, if the stellar mass of each galaxy is computed independently, the total inferred $f_{\star}$ is significantly lower than that reported by previous works which assume a constant $M^*/L$. \citet{lea12} attribute this discrepancy to a combination of the shortfall of the constant $M^*/L$ approximation (with bias towards early-type galaxies, relevant for all mass retrieval methods) and systematic differences in SPS modelling.

Any single $M/L$ ratio assumption inevitably fails to capture the full complexities of a cluster system. Indeed, \citet{lin12} note that a constant $M^*/L$ ratio is robust only in the event of a weak dependence on stellar mass or morphology. In this work, we will assess the validity of the assumption of a constant $M^*/L$ on cluster scales, and the effects of different SPS packages on the inferred cluster-wide value of $M^*/L$.  To this end, we have assembled a sample of nearby clusters with available multi-band optical and X-ray data. We compile cluster stellar luminosities and masses on a galaxy-by-galaxy basis using our own membership assignment scheme. The stellar masses are evaluated using various colour-based mass-to-light ratio transformations, as well as SED-fits to broadband optical photometry, \emph{for each galaxy in the cluster}.  This analysis yields estimates for the total stellar content in these clusters (with limitations, as we will discuss below), as well as cluster-wide stellar mass-to-light ratios, whose variations among clusters is of special interest. This effectively extends the analysis by \citet{lea12} to halos in the mass range $10^{14}M_\sun$ to $10^{15}M_\sun$.  

Our paper is organised as follows: the cluster sample and data source are described in Section 2 and our treatment of the cluster membership -- a chief source of uncertainty in the absence of full spectroscopic coverage -- is presented in \se{method}.  Stellar mass derivations and the convergence to our preferred approach are addressed in \se{stellarmass}.  Results and discussions are found in \se{results}. Our main conclusions and a look towards future investigations are presented in \se{conclusions}. 

Throughout this work, we adopt the $\Lambda$CDM cosmology with $H_0=71$ km s$^{-1}$ Mpc$^{-1}$, $\Omega_M=0.27$, and $\Omega_{\Lambda}=0.73$.


\section{Sample and Data}\label{sec:sampdata}

\subsection{Sample} \label{sec:sample}

Our proposed investigation requires reliable X-ray measurements, which constrain the total halo mass (assuming hydrostatic equilibrium), and multi-band optical photometry, which will facilitate the computation of galaxy stellar masses. We draw from three published X-ray studies of nearby galaxy clusters, all based on {\emph{Chandra}} observations: \citet{vik06}'s first sample of nearby relaxed clusters, \citet{vik09}'s expanded sample as a follow-up to the {\emph{ROSAT}} $400deg^2$ survey, and \citet{sun09}'s analysis of low-mass clusters and groups. We require that the selected clusters overlap with the Sloan Digital Sky Survey (SDSS) Data Release 10 \citep[DR10;][]{ahn13} which provides optical photometry ({\emph{u,g,r,i,z}}) and spectroscopy for a flux-limited sample of galaxies within each cluster. We impose a redshift range of $0.04<z<0.1$ in order to exclude both very nearby clusters (e.g. Virgo), for which ``shredding'' of large galaxies is an issue, and distant clusters, for which SDSS fails to sample the cluster galaxy luminosity function to sufficient depth. Additional selection criteria include dynamic relaxedness (based on visual inspection of X-ray morphology), availability of total mass estimates (M$_{500}$ from X-ray spectroscopy, assuming hydrostatic equilibrium), and some spectroscopic coverage (to anchor bright cluster members). 

The final sample of 12 clusters, tabulated in \Table{VikSample}, spans a range in halo mass of $0.5-12\times10^{14}M_\sun$. The virial mass estimates, $M_{500}$, are taken from the literature (see \Table{VikSample}) and represent the total mass enclosed within the radius $R_{500}$ \footnote{$r_{\Delta}$ is the radius within which the mean enclosed density is $\Delta * \rho_{crit}$, provisionally defining the outer boundary of a cluster.}. Given $M_{500}$ and $\rho_{crit}$ at the cluster redshift, $R_{500}$ can be computed from: 

\be
R_{500}=\left(\frac{3}{4 \pi}\frac{M_{500}}{500 \rho_{crit}}\right)^{1/3}
\label{rdelta}
\ee

\begin{table}
\begin{center}
\caption{Properties of galaxy clusters in our sample.}
\begin{tabular}{cccccc}
\hline\hline
Cluster & RA  & Dec  & z & M$_{500}$ & Ref\\
 & ($^{\circ}$) & ($^{\circ}$) & & (10$^{14}$ M$_{\odot}$) & \\
\hline
A 85 & 10.460 & -9.303 & 0.0557 & 5.98 &1 \\ 
A 160 & 18.248 & 15.491 & 0.0447 & 0.79 & 3 \\
A 1650 & 194.673  & -1.761 & 0.0823 & 4.59  & 1 \\
A 1692 & 198.057 & -0.974 & 0.0848 & 0.970 & 3 \\
A 1795 & 207.217 & 26.591 &  0.0622 & 5.46 & 1 \\
A 1991 & 223.631 & 18.642 & 0.0592 & 1.23 & 2\\
A 2029 & 227.734 & 5.745 & 0.0779 & 8.01 & 2 \\ 
A 2142 & 239.583 & 27.233 & 0.0904 & 11.96 & 1 \\
A 2244 & 255.677 & 34.060 & 0.0989 & 5.11 & 1 \\
MKW 3s & 230.466 & 7.709 & 0.0453 & 2.09 & 1 \\
UGC 842 & 19.723 & -1.002 & 0.0452 & 0.560 & 3 \\ 
Zw1215 & 184.421 & 3.656 & 0.0767 & 5.75 & 1 \\
\hline \\
\end{tabular}
\end{center}
\vspace{-0.2in}
1:~\cite{vik06}; 2:~\cite{vik09}; \\3:~\cite{sun09}
\label{VikSample}
\end{table}

\subsection{Data}\label{sec:data}

Optical photometry and, where possible, spectroscopic redshifts, are obtained from SDSS DR10 for all galaxies in each cluster that lie within an angular separation corresponding to  $R_{500}$ on the sky. Throughout this work we utilize SDSS's ``composite model magnitudes'' (cmodelMag), which represent a linear combination of the best-fit exponential and $r^{1/4}$ \citep{devauc48} models to the galaxy surface brightness profile \citep{sto02}. 

We caution that SDSS model magnitudes and sizes may still be subject to systematic errors, such as deprojection effects \citep[e.g.,][]{hall12}, a detailed investigation of which is beyond the scope of the present study. Notably, several recent studies \citep[e.g.,][ and references therein]{ber13} have shown that SDSS model photometry underpredicts luminosity from the brightest galaxies. 
We do not expect our conclusions to be affected by such a deficiency in the overall stellar content, assuming that the stellar populations (specifically the stellar mass-to-light ratio) do not vary in the galaxies' outskirts (which we may be missing). Based on a careful re-analysis of SDSS imaging for Virgo cluster galaxies, Roediger et al. (2011) find that the stellar populations gradients at large galactocentric radii are flat.  Nevertheless, we present an analysis of possible photometric biases and their effect on our study in Section 5.3.
Galactic extinction estimates for each galaxy are obtained from SDSS and are used to compute extinction-corrected fluxes. K-correction terms provided by SDSS will be used for CMLR-based stellar mass estimates (see \se{stellarmass}).


\section{Methodology}\label{sec:method}

We now describe the method by which the total luminosity and stellar mass for each cluster are derived.

\begin{figure*}[htb]
\center
\includegraphics[width=0.95\textwidth,trim=2.2cm 14.5cm 2.2cm 1.5cm]{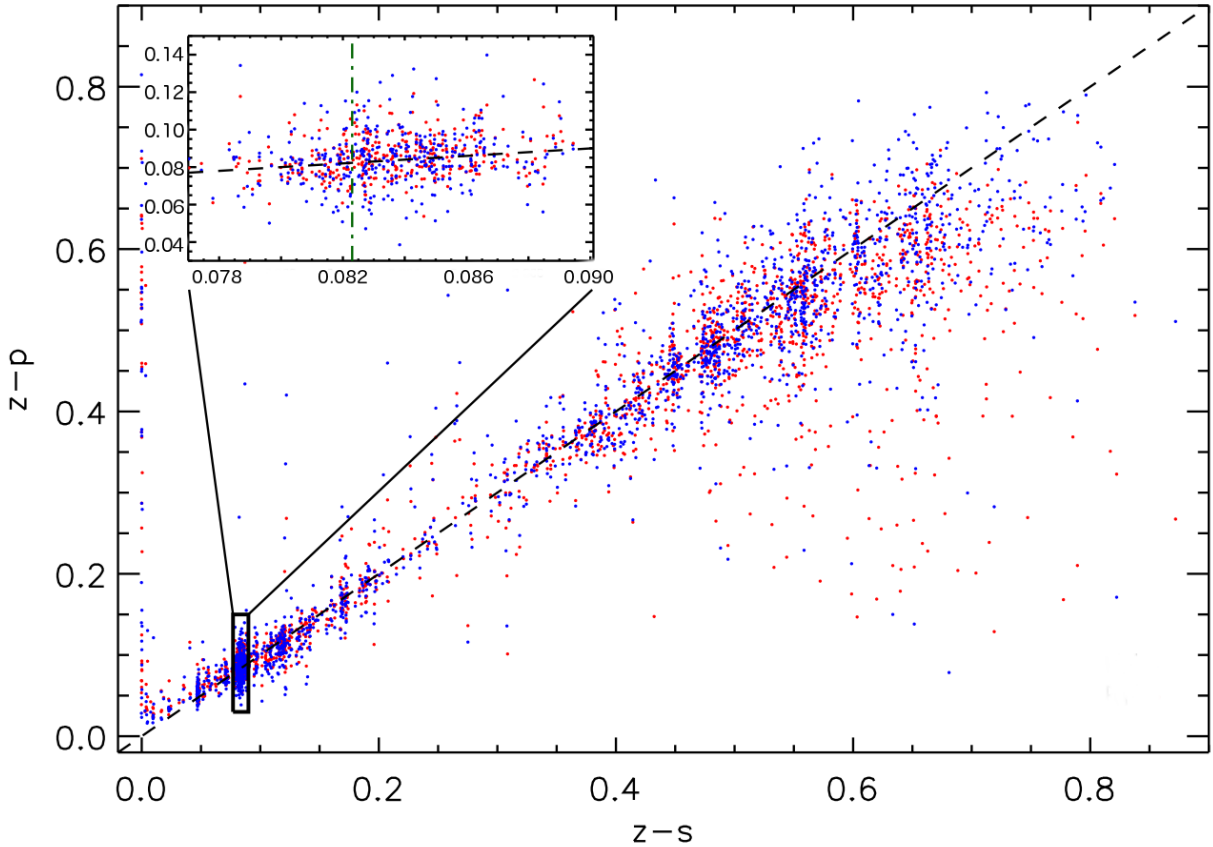}
\caption{Comparison between $z_{phot}$ (z-p) as deduced by methods KF (blue dots) and RF (red dots), versus $z_{spec}$ (z-s), for galaxies with SDSS spectroscopy around the position of A1650. The dashed diagonal represents equality between $z_{phot}$ and $z_{spec}$. The inset magnifies the heavily populated region around the cluster redshift. Overall, KF and RF photometric redshifts have similar accuracy and scatter properties, especially over redshift range of relevance for this study.}
\label{zpzs-KFvsRF-1650}
\end{figure*}

In order to accurately measure the total luminosity and stellar mass for a given galaxy cluster, the identification of all cluster galaxies is crucial. Ideally, we wish to maximize completeness in cluster member detections whilst minimizing contamination from foreground and/or background galaxies due to projection. Traditionally, cluster-searching algorithms have been plagued by the trade-off between these two requirements. Whereas the most conservative membership algorithm, which only considers spectroscopically-confirmed galaxies, achieves minimal contamination, it is observationally expensive and requires pre-selection of spectroscopic targets which may be biased towards bright, red sequence galaxies.

Conversely, purely photometric membership assessment (e.g. selection by colour through red-sequence identifications or by a luminosity cut) assume a strict uniformity in the galaxy population and can miss blue cluster members (i.e. incompleteness), or incorrectly include field galaxies with fortuitous colours that match the cluster's average (i.e. contamination). 
Photometric redshifts, typically based on SED-fitting \citep[see e.g.,][]{bolzonella00} allow for more robust rejection of foreground/background contamination; however, the achieved precision is not sufficient for definitive individual member assignments. As a compromise, we employ spectroscopic redshifts where available (see \Table{spect-perct}) and utilize photometric redshifts as distance indicators in a statistical sense in order to reject foreground and background galaxy populations without imparting a colour bias on our galaxy sample.

We require that all identified galaxies be detected in both binned and unbinned data, be appropriately treated for blending, and have no major photometric issues\footnote{SDSS parameters are used to perform photometric quality checks and derive fluxes as follows: for each galaxy to be considered, cmodelMag must be defined within each band, and z-phot must exist. We demand that ``other'' flags not be present to maintain the photometric integrity of our galaxies, except where possibly the BCG or a spectroscopically confirmed member is involved. These ``other'' flags are, following  \citealt{sza11}: TOO\_FEW\_GOOD\_DETECTIONS, NOTCHECKED\_CENTER, NOPROFILE, and BADSKY.}. We adopt a brightness limit of $r<22$, below which SDSS is generally incomplete, and require that all galaxies lie within a distance of $R_{500}$ from the BCG.

We use photometric redshifts ($z_{phot}$) from the photo-z table in the SDSS3 DR10 which is defined for all galaxies with reliable photometry in the survey. These are derived from five-band photometry using a KD-tree nearest neighbour fit (KF), as described in \citet{csa07}.  The alternative $z_{phot}$ catalogue in DR10 (PhotozRF), uses random forests to determine these values \citep{car10}. Since the two techniques yield similar accuracy and statistical properties (see \Fig{zpzs-KFvsRF-1650}), we use the former, which also contains the quality assurance parameter $nnCount$ to indicate the extent of the spectroscopic training set coverage. K-correction in each waveband is deduced via SED template fitting as part of DR10. 

For each cluster in our sample, we generate 1000 realizations, each containing a subset of all galaxies found within the projected $R_{500}$. The likelihood of a galaxy being included in any given cluster realization depends on the probability of its measured $z$ falling within the cluster range, taken to be $z_{clust} \pm \delta z$, where $\delta z$ encompasses the expected velocity dispersion for a given cluster characteristic radius. $\sigma_v$ is related to $R_{500}$ via the relation from \citealt{zha11a}:
\be
log\left(\frac{R_{500}}{kpc}\right) = 3.07 + 0.89 ~log\left(\frac{\sigma_v}{1000~km s^{-1}}\right)
\label{r500-to-sigmav}
\ee
We approximate the acceptable range of redshifts for galaxies bound to the cluster as $\delta z \approx dv/c  = 2\sigma_v/c$, thus allowing galaxies with line-of-sight velocities within $dv=2\sigma_v$ of the cluster centre. 
Column 3 of \Table{SampleSize} provides $\delta z$ for each cluster.  

The adoption of a radial velocity interval bracketing the cluster's cosmological redshift as a criterion for assigning cluster membership differs from exclusively selecting galaxies located within the sphere of radius $R_{500}$ around the cluster centre. In particular, infalling galaxies outside the nominal cluster search radius, as well as foreground/background galaxies, could spoil this operation. We assess these effects below using simulated clusters from the Millennium Simulation \citep{springel05}.

Assuming that the probability distributions for both $z_{phot}$ and $z_{spec}$ are Gaussian, with widths given by the quoted redshift uncertainty ($z_{err}$), the probability that a given galaxy belongs in the cluster can be:
\be
P~(member)=\int_{z_{cl}-\delta z}^{z_{cl}+\delta z}\frac{1}{\sqrt{2\pi}z_{err}} exp\left(-\frac{(z-z_{gal})^2}{2z_{err}^2}\right) dz.
\label{gauss-zp}
\ee

\noindent{}
In practice, a random $z_{gal}$ value is drawn from the Gaussian distribution as in the integrand of \Eq{gauss-zp} and compared to $z_{clust}$. If it falls within the interval $z_{clust} \pm \delta z$, then it is accepted. Otherwise, it is excluded from any particular cluster realization. A library of 1000 such realizations is produced for each cluster, thus statistically accounting for the membership uncertainty due to uncertainty in distance. This action offsets contamination while still considering every galaxy detected within the projected cluster radius. The distributions of luminosity, mass, and $M^*/L$ for the 1000 realizations provide the relevant uncertainty estimates.  Given the order-of-magnitude precision improvement on spectroscopic redshifts, clusters with many spectroscopically-confirmed members will naturally have more tightly constrained measurements. The number of photometrically- and spectroscopically-confirmed galaxies in each cluster are summarized in \Table{spect-perct}.

\begin{table}
\begin{center}
  \caption{Estimated metrics for the cluster sample.}
 \label{SampleSize}
\begin{tabular}{ccccccc}
\hline\hline
Cluster & $z$ & $\delta z^1$ & M$_{500}$ & M$_{200}$$^2$ & r$_{500}$ & r$_{200}$$^2$\\
 & & & \multicolumn{2}{c}{(10$^{14}$ M$_{\odot}$)} & \multicolumn{2}{c}{(Mpc)} \\
 \hline
 A 85 & 0.0557 &  0.007 & 5.98 & 11.49 & 1.249   & 2.11 \\ 
A 160 & 0.0447 & 0.003 & 0.79 & 1.43 & 0.626 & 1.05 \\
A 1650 & 0.0823 &  0.006 & 4.59  & 8.58 & 1.135 & 1.90  \\
A 1692 & 0.0848 & 0.003 & 0.970 & 1.72 & 0.658 & 1.11 \\
A 1795 & 0.0622 & 0.007 & 5.46 & 10.41 & 1.235  & 2.04 \\
A 1991 & 0.0592 & 0.004 & 1.23 & 2.23 & 0.732 & 1.23 \\
A 2029 & 0.0779 & 0.008 & 8.01 & 15.3 & 1.362 & 2.31 \\
A 2142 & 0.0904 & 0.009 & 11.96 & 23.03 & 1.558 & 2.63 \\
A 2244 & 0.0989 &  0.007 & 5.11 & 9.49 & 1.170 & 1.95 \\
MKW 3s & 0.0453 & 0.005 & 2.09 & 3.89 & 0.882 & 1.47 \\
UGC 842 & 0.0452 & 0.003 & 0.560 & 0.99 & 0.570 & 9.36 \\ 
Zw1215 & 0.0767 & 0.007 &  5.75 & 10.88 & 1.225 &  2.06  \\
 \hline
 \end{tabular}
 \end{center}
\vspace{-0.05 in}

 $^1$: Velocity width of cluster from \cite{zha11}.\\
 $^2$: Assumes NFW dark matter halo \citep{duf08}.

\end{table}

\begin{table}
\begin{center}
\caption{Number of galaxies detected in each cluster.}
\label{spect-perct}
\begin{tabular}{ccccccc}
\hline\hline
Cluster & N$^{tot}_{gal}$ ($r<R_{500}$) & w/ spec & N$^{spec}_{gal}$ & N$^{boot}_{gal}$ \\
\hline
A 85  & 2640 & 140 & 87 & 167 \\
A 160  & 856 & 67 & 33 & 34 \\
A 1650  & 1183 & 64 & 28 & 79 \\
A 1692 & 303 & 20 & 16  & 19 \\
A 1795  & 2408 & 145 & 82 & 144 \\
A 1991  & 1086 & 70 & 35 & 51 \\
A 2029 & 1809 & 149 & 94 &  173 \\
A 2142  & 2195 & 193 & 102 & 254 \\
A 2244  & 1226 & 91 & 48 & 108 \\
MKW 3s  & 1627 & 97 & 45 & 68 \\
UGC 842 & 652 & 65 & 26  & 29 \\ 
Zw1215 & 1330 & 127 & 66 & 141  \\
\hline
\end{tabular}
\end{center}
\vspace{-0.05 in}

N$^{tot}_{gal}$: Number of photometrically-detected galaxies within $R_{500}$.\\
N$^{spec}_{gal}$: Number of spectroscopically confirmed cluster galaxies.\\
N$^{boot}_{gal}$: Median number of cluster galaxies from bootstrap analysis.

\end{table}

\begin{figure}[htb]
\center
\includegraphics[width=0.48\textwidth]{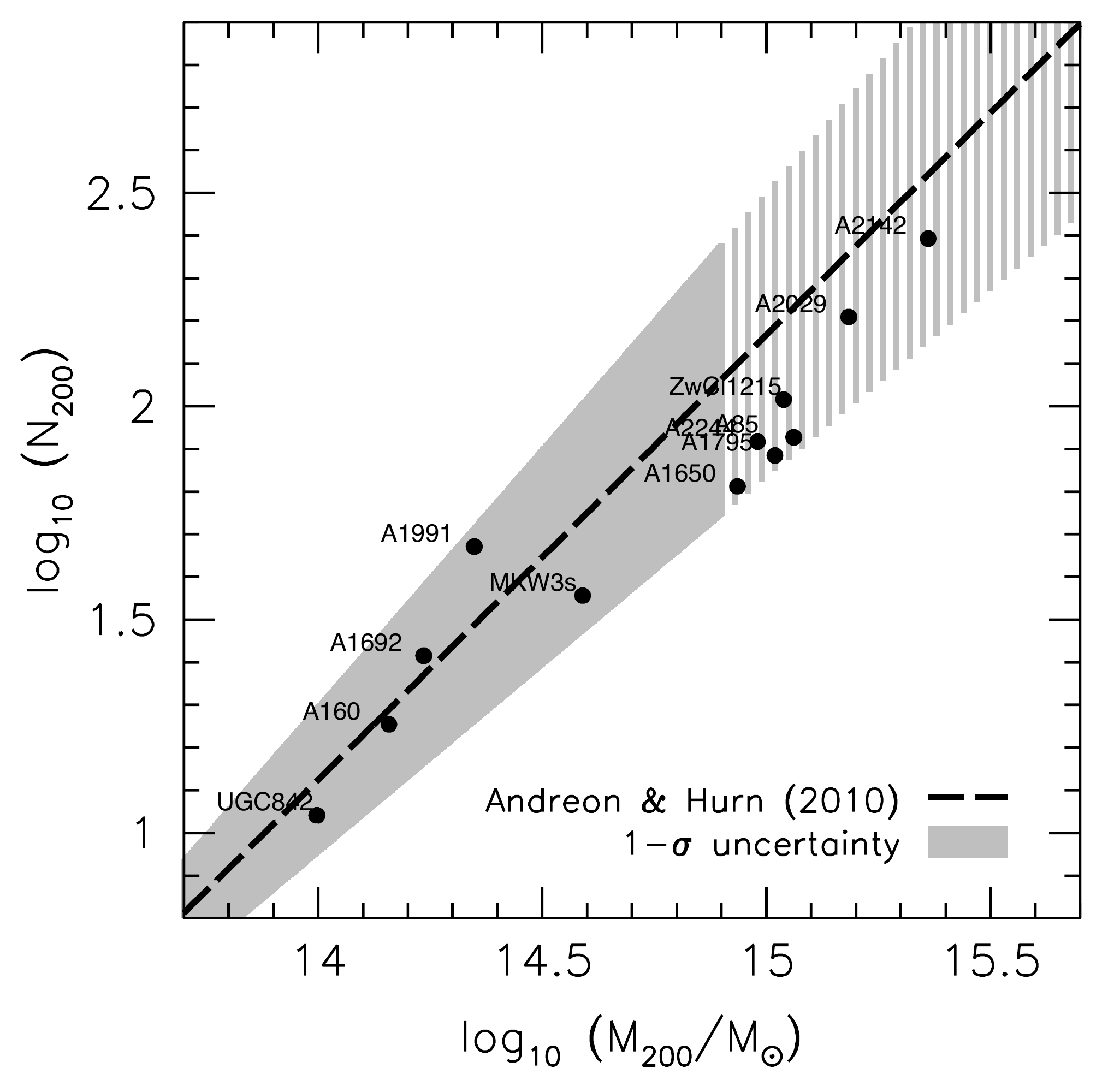}
\caption{Comparison between the $M_{200}$ and $N_{200}$ scaling relation of \citet{and-hu10} and that from our $z_{phot}$-based Monte Carlo cluster construction. The shaded area represents the combined $1\sigma$ slope uncertainty and scatter quoted by Andreon and Hurn. Note that this relation is only well-constrained for $13.7 < \log_{10}$M$_{200} < 14.9$ -- we have highlighted the fact that we are showing an extrapolation by using a hashed fill for $\log_{10}$M$_{200} > 14.9$.}
\label{and-richness-mass200}
\end{figure}

\begin{figure}[htb]
\center
\vspace{-180pt}
\includegraphics[width=0.48\textwidth,trim=2.2cm 2cm 2.2cm 0cm]{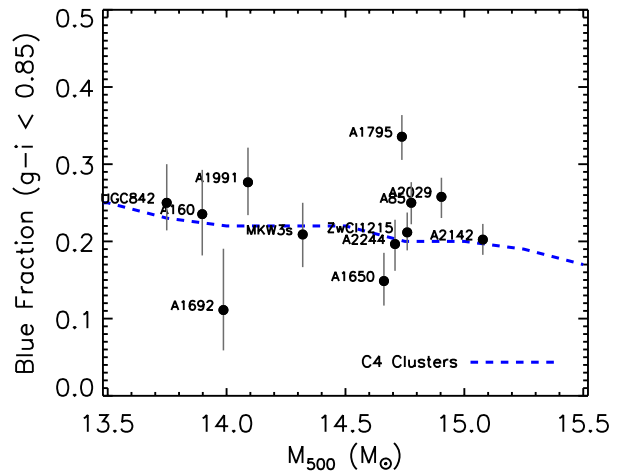}
\caption{Comparison of the fraction of blue galaxies (with $g-i <$ 0.85) from our analysis to the typical cluster spiral fraction as a function of halo mass. The dashed line follows the occurrence of spiral galaxies in the C4 Clusters sample \citep{mil05} as detected by \citet{hoy12} via use of the Galaxy Zoo project. While C4's downward trend is not apparently replicated by our (limited) sample, the overall agreement is satisfactory given the large uncertainties in both measurements.}
\label{bluefract}
\end{figure}

To assess the robustness of our $z_{phot}$-based cluster construction scheme, we perform several sanity checks. For one, we can compare our measured mass--richness relation with previous works. \cite{and-hu10}, who define richness ($N_{200}$) as the number of bright ($M_V < -20$) red galaxies within $R_{200}$, find:
\be
log\left(\frac{M_{200}}{M_\sun}\right)=(0.96\pm0.15)(log(N_{200})-1.5)+(14.36\pm0.04),
\label{and-rich-m200}
\ee
from a fit to 53 cluster caustic masses in a Bayesian framework. Note the Andreon \& Hurn study only involves clusters with $M_{200} \lesssim 10^{15} M_\sun$ (see their Figure 2).

To ensure a fair comparison, we first convert from $R_{500}, M_{500}$ to $R_{200}, M_{200}$ assuming an NFW profile for the cluster halo distribution, following \citet{duf08} (see \Table{SampleSize}). Adopting $R_{200}$ also modifies the search radius around each cluster for member galaxies, requiring us to re-run the cluster-construction bootstrap analysis after first discarding all galaxies that fall below the $M_V < -20$ brightness criterion of \cite{and-hu10} \citep[assuming $V = g - 0.5784(g - r) - 0.0038$; ][]{lup05}. 

\Fig{and-richness-mass200} summarizes the results of this comparison. Our cluster points, built from the described $z_{phot}$-PDF based member selection described above, agree well with both the slope and scatter quoted by \cite{and-hu10} in the lower cluster mass ranges (i.e. $\lesssim10^{15}$ M$_\sun$) where the two samples overlap. At $\gtrsim10^{15}$ M$_{\sun}$, we find systematically lower values of N$_{200}$ compared to \cite{and-hu10}, though the two samples still agree within the 1$\sigma$ uncertainties (see shaded region in \Fig{and-richness-mass200}). We measure a slightly shallower slope to the N$_{200}$--M$_{200}$ relation of $0.82 \pm 0.08$, consistent at the 1$\sigma$ level with the value of $1.04 \pm 0.16$ quoted by \cite{and-hu10}.

We further examine whether the measured fractions of blue galaxies in the clusters are realistic. Blue galaxies are defined here as having extinction-corrected $g-i < 0.85$. \Fig{bluefract} illustrates, as a function of $M_{500}$, the median and $1\sigma$ fraction of blue galaxies with respect to total number of cluster members over 1000 realizations. Overplotted are results from measurements of spiral fractions in the C4 cluster catalogue \citep{mil05} by \citet{hoy12}. Although the two metrics are not identical, they are expected to be similar. Our measured blue fractions ($\sim$20--30\%) are consistent with the expected spiral fraction in massive clusters ($10^{13.5} M_{\odot} < M_{500} < 10^{15} M_{\odot}$), further validating our method.

\begin{figure}[htb]
\center
\vspace{-10pt}
\includegraphics[width=0.48\textwidth]{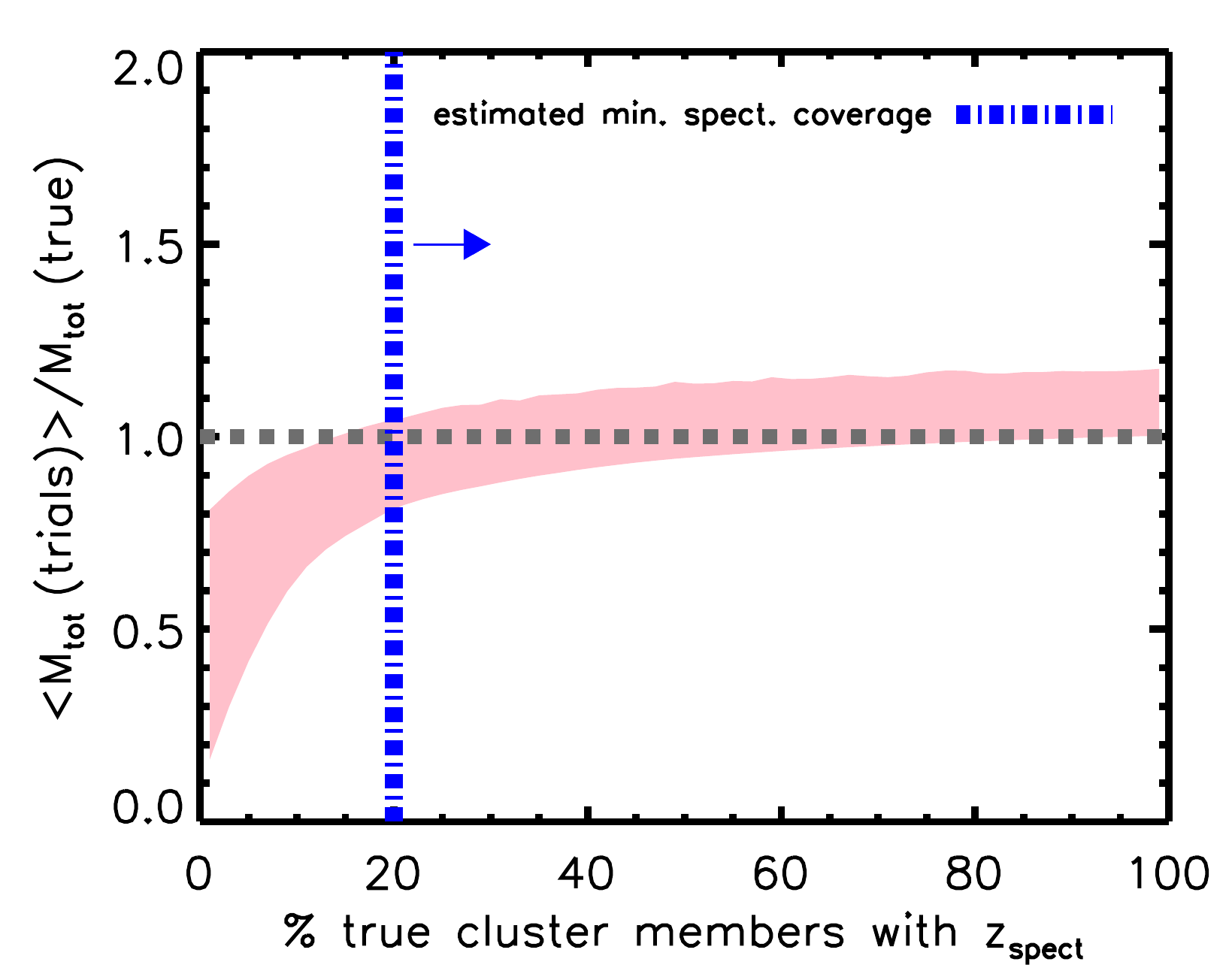}
\caption{Demonstration of the accuracy of aggregate mass estimates from the zphot bootstrap method described in Section 3, using galaxy clusters in the Millenium Simulation where the ``true'' mass is known.  The probabilistic total mass in galaxy halos obtained from cluster realizations is compared with the (fixed and known) total mass in true members.  The shaded region spans the range of outcomes corresponding to each cluster/orientation combination.  See text for more details about the simulations.  For clusters with $>20$\% spectroscopic member confirmation - satisfied for all the clusters in this study - the derived halo masses are at most 20\% (typically $<10$\%) discrepant from their true value.}
\label{mme_msim_contam}
\end{figure}

As another important check, we gauge the discrepancy between the actual aggregate mass (i.e. from all true cluster members within the spherical volume defined by $R_{500}$) and that inferred by our $z_{phot}$ probabilistic scheme. We do this using simulated clusters and field galaxy halos from the Millenium Simulation \citep{springel05}. We choose two group-sized halos ($\sim$50 members) and two cluster-sized halos ($\sim$300 members) at $z\sim0.06$. These structures are identified using a friends-of-friends algorithm. Individual galaxy properties, including stellar mass, for each subhalo are provided by \cite{delucia06}. For each group/cluster, we extract three-dimensional positions and stellar masses for all galaxies within $R_{200,crit}$ and with M$_* > 10^9$ M$_{\odot}$ (roughly mimicking our sensitivity limit), projecting the cluster on the sky along three different axes for each cluster (probing different line-of-sight structure). For each cluster and line-of-sight orientation, we obtain the true redshift of every galaxy halo within the cylinder of $R_{200,crit}$ (converting line-of-sight distance to radial velocity) and assign to them a characteristic photometric redshift uncertainty $dz_{phot}=0.03$. This choice is motivated by typical errors on $z_{phot}$ as reported in the SDSS photo-z table for galaxies at $z=0-0.2$ (see also Fig.\ 1). For a subset of the true cluster members we also assign spectroscopic redshifts -- that is, PDFs which have negligible error centered on the true $z$ -- in order to quantify the effects of spectroscopic incompleteness.

We then perform a bootstrap analysis following the procedure described above (assuming a Gaussian PDF with width $dz_{phot}=0.03$ for $z_{phot}$) to build cluster realizations, and tabulate the resultant total stellar masses. As a function of the fractional spectroscopic coverage for true cluster members (assuming more massive cluster member are prioritized in spectra acquisition, which is typically the case for SDSS), the statistically inferred and true masses are, on average, within 20\% of one another once we have at least 10\% spectroscopic confirmation (\Fig{mme_msim_contam}). While we do not know the ``true'' number of galaxies within a three-dimensional radius for the clusters in our sample, we estimate from richness-halo mass relations -- calibrated on well-studied, low-redshift clusters (\citealt{and-hu10}; \citealt{high10}) -- that our typical spectroscopic completeness is $>$20\% for the systems analyzed in this study.

\Fig{mme_msim_contam} illustrates the bias in the total stellar mass for the 12 cluster-line-of-sight combinations simulated. Not surprisingly, sparse spectroscopic coverage can cause significant downward bias in the inferred mass metric for the cluster, due to cluster members with large redshift uncertainties scattering out of the cluster volume. Fortunately, in this regime the marginal rate of improvement with spectroscopic redshifts is also high. 
Again, if $z_{spect}$ is available for 10\% or more of the most prominent cluster galaxies, the typical error on galaxy mass derived from our method is less than 20\%. Note that this discrepancy incorporates contributions from the distinction between cylinder and sphere in our 3D volume selection.
The positive bias in the regime of high-completeness spectroscopic coverage is due to contamination of foreground/background galaxies and the geometric factor between spherical and cylindrical volume selection. This bias can be reduced with some spectroscopic coverage of the field population, potentially excluding heavy-weight contaminants. While such $z_{spect}$ for field galaxies is not considered in the present test, it has been used for membership exclusion in the analysis of our sample. 
This implies a $\sim$10\% uncertainty on the true stellar mass for systems studied in this paper, which is less than the uncertainty in the X-ray-derived total cluster mass. 

The measurement of the blue fraction will be similarly biased.  Using the same techniques outlined above, we find that the blue fraction will be biased high by $\sim$8\% for typical clusters in our sample, due to field galaxies lying along the line of sight. This bias is small -- our typical measurement uncertainty is $\sim$5--10\% -- though it may be large enough to wash out any subtle trends with blue fraction, since the overall range in blue fraction only spans $\sim$25\% in this sample (see \Fig{bluefract}).

We caution that the scheme thus far outlined has obvious limitations. First, the aforementioned probabilistic treatment of galaxy membership, as in \Eq{gauss-zp}, is oversimplified. 
Ideally, one would marginalize over the precise $z_{phot}$ PDFs in favour of the fiducial Gaussian form employed here. Such avenues are beyond the scope of the current paper.
Second, our Monte Carlo method is naturally biased to galaxies with spectroscopic redshifts. Spectroscopic redshifts have very sharply peaked PDFs with error widths of $\sim 10^{-4}$, and thus tend to be included in either 0\% or 100\% of the cluster realizations. 
In the test above, we have demonstrated the importance of at least modest spectroscopic coverage in securing the validity of our methodology, but with a specific assumption about the availability of spectra as a function of cluster member properties (i.e. preference is given to more massive galaxies). In reality, whether a galaxy is part of a spectroscopic target program is somewhat arbitrary (although spectroscopic coverage is certainly skewed towards the brightest members), leading to a somewhat asymmetric selection bias which entails a greater uncertainty than is captured in the PDFs alone.

Other errors inherent to our cluster galaxy stellar mass accounting technique include incompleteness due to the magnitude-limited nature of the survey sample, as well as deprojection effects due to the selection of galaxies within a redshift cylinder as opposed to a sphere. The former problem biases our resultant total stellar mass low, while the latter overestimates the total number of galaxies in each cluster. A correction factor could be applied to compensate for each effect as gauged by, e.g., mock catalogs. A similar treatment by \citet{lea12} for COSMOS galaxy groups yielded a correction factor $<$15\% which is estimated to be much less than systematic errors in the stellar mass estimates themselves (\se{results}). 
Such a correction is also consistent with \Fig{mme_msim_contam}.
As described above, we find from halo-only simulations that the mass bias due to selection on radial velocity (rather than a three-dimensional cut) is $\sim$10\% for a wide range in cluster masses. We therefore ignore this effect for the present study. As long as the faintest, low-mass systems, as well as those residing in the space between the sphere and its cylindrical counterpart are not strongly skewing the $M^*/L$ distribution, then the measured $M^*/L$ should be representative of the true value. We briefly return to this point in \se{uncertain}.

\section{Stellar Mass Estimation}\label{sec:stellarmass}

As discussed in \se{data}, stellar masses of galaxies are popularly inferred from photometry via simple colour-M*/L (CMLR) (fast, but with large uncertainties), or full-scale SED-fitting (more reliable but expensive) on which the CMLRs are based.  Below we explore both of these approaches, providing multiple estimates of stellar mass for each cluster and allowing us to assess the similarities and differences of each formalism. 

\subsection{Stellar Masses via Colour-M*/L Transformations}

Numerous CMLRs are now available, some based purely on stellar evolution model libraries (e.g. \citet{zib09} [Zi09], \citet{ip13} [Ip13]), and others depending on a subset of these libraries constrained by observation (e.g. \citet{bel03} [Be03], \citet{tay11} [Ta11]). These relations are generically expressed as:

\be
log_{10}\left(\frac{M^*}{L_{\lambda}}\right) = a_{\lambda, colour} + (b_{\lambda, colour}\times colour) ,
\label{colourml-eqn}
\ee

\noindent{}where $a_{\lambda,colour}$ and $b_{\lambda,colour}$ are the normalization and slope, respectively, for a given relation. Each of the aforementioned papers provide waveband- and colour-dependent $a_{\lambda,colour}$ and $b_{\lambda,colour}$ terms, which we summarize in \Table{cmlrabs}. 
The amount of scatter in a given CMLR relation varies with both the luminosity band and colour chosen. For this work, we adopt the $i$-band luminosity with the $g-i$ color due to their higher S/N ratio over other SDSS bands, broad baseline, and stable CMLR \citep{tay11}.

The variance in the slope and the normalization of these relations highlights large systematic differences between the relations. In general, the large differences in $a_{i,g-i}$ from \Table{cmlrabs} arise from different choices of IMF between authors. Further discrepancies in both $a_{i,g-i}$ and $b_{i,g-i}$ between authors arise from a variety of assumptions implicit in the stellar population libraries (SPL), including details of stellar evolution models (e.g. AGB modelling), star formation history, and treatment of dust extinction.
Each CMLR also depends on the empirical training dataset used to refine the relations. \Table{cmlr-origins} summarizes some of these model assumptions for several popular CMLRs. For a recent review of colour diagnostics for galaxy $M^*/L$'s, see \citet{courteau14}. 

\begin{table}
\centering
\caption{Coefficients for the relation $log_{10}(M*/L_{\lambda}) = a_{\lambda, col} + (b_{\lambda, col}*col)$}
\begin{tabular}{ccc}

\hline\hline
Reference & $a_{i,g-i}$ & $b_{i,g-i}$ \\
\hline
\citet{bel03} diet Salpeter & -0.152 & 0.518 \\
\citet{bel03} Kroupa or Kennicutt & -0.302 & 0.518 \\
\citet{zib09} & -0.963 & 1.032 \\
\citet{tay11} & -0.68   & 0.70 \\ 
\citet{ip13}   & -0.625 & 0.897  \\
\hline
\end{tabular}
\label{cmlrabs}
\end{table}

\begin{table*}[htb]
\begin{center}
\caption{Properties of various stellar population models from the literature.}
\label{cmlr-origins}
\begin{tabular}{ccccccc}
\hline\hline
Ref. & SPS & IMF & SFH & Metallicity & Dust & Test Dataset\\
\hline
Be03 &  P\'EGASE \tablenotemark{a} & diet Salpeter \tablenotemark{b} & exponential SFR \tablenotemark{c} & -- & n/a & SDSS EDR \tablenotemark{d} \\
 &  & Kroupa or Kennicutt \tablenotemark{e} &  &  &  &  \\ 
Zi09 & CB07 \tablenotemark{f} & Chabrier \tablenotemark{g} & exponential + bursts & 0.2 to 2 solar & CF00 \tablenotemark{h} & n/a \\
Ta11 & BC03 \tablenotemark{i} & Chabrier & exponential & 0.005 to 2.5 solar & Calzetti \tablenotemark{j} & GAMA \tablenotemark{k} \\ 
Ip13 & Padova Isochrones \tablenotemark{l} &  Kroupa &  exponential + $b$-parameter \tablenotemark{m} & 0.2 to 1.5 solar & n/a & n/a \\
MAGPHYS & CB07 & Chabrier & exponential + bursts & solar &  -- & n/a \\
\hline
\end{tabular}
\end{center}
$^a$: P\'EGASE models \citep{fio-roc97}; $^b$ diet Salpeter IMF \citep{beldJ01}; $^c$: generically, $SFR(t) \propto e^{-t/\tau}$; $^d$: Early Data Release \citep{sto02}; $^e$: Kroupa \citep{kro01} or Kennicutt \citep{ken83} IMF; $^f$: CB07 SPS codes \citep{b07}; $^g$: Chabrier IMF \citep{cha03}; $^h$: an angle-averaged model \citep{cha-fall00}; $^i$: BC03 SPS codes \citep{bc03}; $^j$: Calzetti dust obscuration law \citep{cal00}; $^k$: \citep{dri09};  $^l$: Padova isochrones with TB-AGB treatment \citep{mar-gir07}; $^m$: birthrate = $\psi(T_{present})/<\psi>$, \citep{ip13}; $^n$: dust around AGB stars.
\end{table*}

For each individual galaxy, we use $i$-band luminosities and $g-i$ colours, corrected for Galactic extinction and k-corrected to $z=0$, to arrive at an M*/L using the various CMLRs described in \Table{cmlrabs}.   Stellar masses are computed via:
\be
M^* = L_{\lambda} \times (M^* / L_{\lambda}),
\label{colour-smass}
\ee
where $L_{\lambda}$ is the $i$-band luminosity.

\subsection{Stellar Masses via SED-Fitting with MAGPHYS}\label{sec:magphys}
 
Although time-consuming and somewhat model dependent, SED-fitting should provide more robust estimates of individual galaxy properties, due to its reliance on all five photometric bands ($u,g,r,i,z$), rather than only two ($g,i$). We use the publicly available software Multi-wavelength Analysis of Galaxy Physical Properties (MAGPHYS), a FORTRAN77 program developed by \citet{dC08}, for our analysis. The SPL in MAGPHYS is constructed from the 2007 version of the Bruzual and Charlot (BC03, CB07) stellar population synthesis code \citep{bc03,b07}, which contains an improved treatment of TB-AGB stars. A Chabrier IMF \citep{cha03} is assumed and dust attenuation follows the \citet{cha-fall00} model. A wide range of SFHs are considered, and the parameter space is surveyed based on the notion that any SFH can be dissected into an underlying continuum of exponentially declining star formation rate (SFR) and a series of bursts \citep[e.g.][]{kauf03}. 

MAGPHYS requires a galaxy redshift ($z_{gal}$) and the specific flux with its error ($f_{\lambda}$, $df_{\lambda}$) in each filter. The fluxes are corrected for Galactic extinction but not k-corrected since the program asssembles its SPL at the galaxy's given redshift, effectively performing its own k-correction. All galaxies that are associated with the cluster for a given realization (see Section \ref{sec:method}) are assigned the same redshift -- this may introduce small systematic effects on a galaxy-by-galaxy basis, but these should disappear for the ensemble if the cluster is assumed to be symmetric along the line of sight.
MAGPHYS outputs the stellar mass PDF for which we report the median value. The error range is indicated by the mass interval between the 16th and 84th percentile.    

\begin{figure}[htb]
\centering
\includegraphics[width=0.48\textwidth,trim=1.4cm 14.5cm 3.2cm 1.5cm]{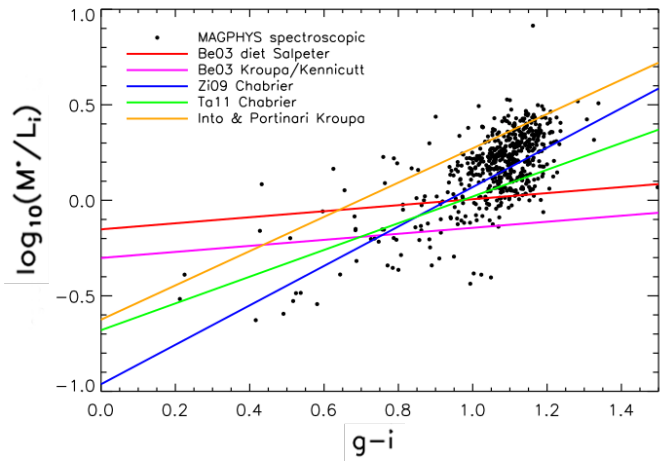}
\caption{$M^*/L_i$ versus ($g-i$) colour for various CMLRs and the MAGPHYS fits to $u,g,r,i,z$ photometry for our spectroscopic subsample (black dots). The spectroscopic sample is preferentially selected for redder (usually brighter) galaxies, so little constraint is placed on the faint/blue regime. Stellar masses derived from SED-fits are matched by the Zi09, Ta11, and Ip13 formalisms and are most consistent with Zi09. Indeed, the underlying model assumptions upon which the Zi09 and MAGPHYS formalism are based (\Table{cmlr-origins}) share similarities. \citet{bel03}, which assumes either a diet Salpeter or Kroupa IMF, predicts systematically higher stellar mass for bluer galaxies, and lower mass for redder galaxies. }
\label{cmlrmp-compare}
\end{figure}

The intrinsic discrepancies between various CMLRs and MAGPHYS are illustrated in \Fig{cmlrmp-compare}. The formulations from \cite{bel03} appear least consistent with the MAGPHYS data and competing formalisms. MAGPHYS-derived values are reasonably well matched by the other CMLRs \citep{zib09, tay11, ip13}.   

\subsection{Boostrapping Aggregation}\label{sec:bootstrap}

We wish to consider the contribution of individual member galaxies to the total stellar mass-to-light ratio. In aggregating the observed luminosity and stellar mass over the detected galaxies, three sources of uncertainty should be explicitly addressed: 1) the measured luminosity (i.e. L-variance), 2) the light-to-mass conversion (i.e. M*-variance), and 3) the membership assessment (i.e. cluster composition variance). 

One approach to this problem involves bootstraping over reasonable PDFs for each quantity, on a galaxy-by-galaxy basis and for each cluster realization, as outlined in \se{method}. Over a large number of simulations, PDFs for total L, M*, and M*/L may be generated and total errors attributable to the aforementioned uncertainty sources are encapsulated in the shapes of these final PDFs.

To this end, we draw, for each member galaxy in a given cluster realization, a random $i$-band luminosity from a Gaussian PDF centered on the extinction-corrected absolute magnitude, with variance as their respective uncertainties. For the CMLR mass conversion, this procedure is repeated for $g$-band luminosity to secure $g-i$. The chosen $i$-luminosity and $g-i$ colour then yield a best-estimate of $log_{10}(M^*/L_{i})$ (i.e \Eq{colourml-eqn}) and \Table{cmlrabs}). Another round of bootstrapping is performed whereby a $\log_{10}(M^*/L_{i})$ is drawn from a Gaussian distribution with centroid at the best-estimate and variance 0.1 dex, which represents the typical uncertainty in the CMLR. Finally, the galaxy's $M^*$ is derived in accordance with \Eq{colour-smass} {\emph{for this particular realization}}.

With MAGPHYS, no colour or direct evaluation of $M^*/L_i$ is involved. Instead, we draw the stellar mass for each galaxy directly from the relevant PDF output by the program, again approximated to be Gaussian. 

For galaxies whose measurement errors on $m_i$ or $m_g$ exceed 0.2, the photometry is deemed unreliable for robust CMLR applications. In such cases ($\lesssim 20\%$ of all candidate galaxies), the stellar mass PDFs are replaced with MAGPHYS values, which takes multiband information and their respective errors into account.      

This process is repeated for 1000 realizations, yielding well-sampled PDFs for $M^*/L_i$ and $M^*$ for each cluster which fully account for our uncertainty in cluster membership, optical photometry, and light-to-mass conversions.

\section{Results \& Discussion}\label{sec:results}

For each CMLR (or SED-fit), stellar mass-to-light ratios, stellar masses, and stellar mass fractions  (relative to $M_{500}$) are extracted from each cluster PDF and compared with cluster luminosity, total halo mass ($M_{500}$), and blue fraction. In this section, we analyse these parameters, compare them with past studies, and interpret them in the context of cluster evolution theories.

\subsection{Stellar Mass-to-Light and Stellar Mass Fraction \\
 vs $i$-band Luminosity, $M_{500}$, and Blue Fraction}\label{sec:masstolight}

\begin{figure*}[p]
\center
\includegraphics[width=0.95\textwidth,trim=0.5cm 1cm 0.5cm 1cm]{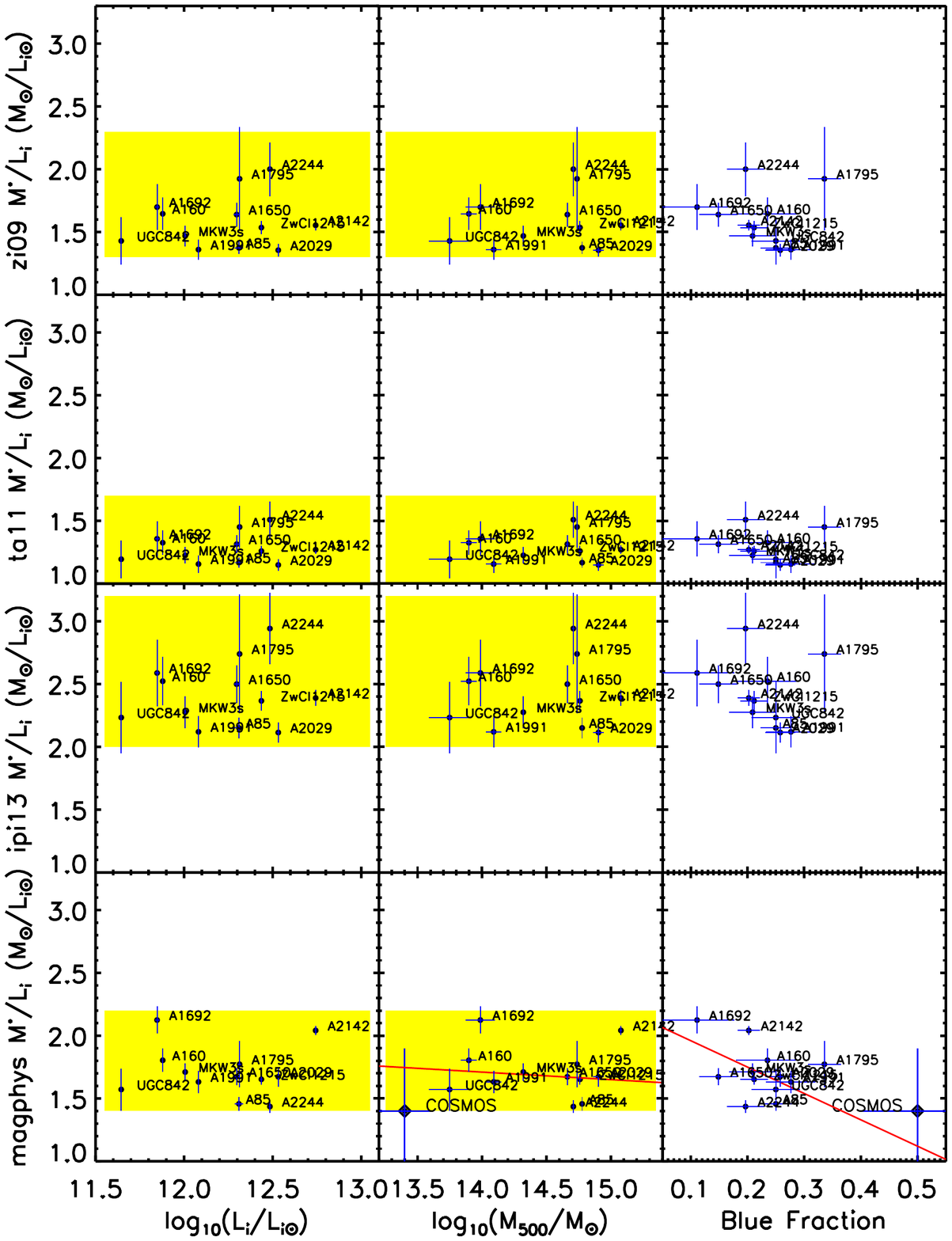}
\vspace{10pt}
\caption{$M^*/L_i$ versus $L_i$, $M_{500}$, and blue fraction for our cluster sample. Shaded regions encompass the range of cluster stellar mass-to-light ratios specific to each stellar mass conversion method. The various methods yield results that are vertically offset from one another.  A statistical analysis shows no significant correlation between $M^*/L_i$ and the total mass (M$_{500}$) or blue fraction of the cluster galaxy population.
Error bars are the statistical from the bootstrap PDFs for each quantity. 
The two red lines represent fits to the data: $M^*/L_i =(-0.06 \pm 0.48) ~ \log_{10}(M_{500}/10^{14}\,M_\sun) + (1.71 \pm 0.23)$ and $M^*/L_i = -(2.10 \pm 2.01) ~ f_{blue} + (2.17 \pm 0.45)$.
}
\label{zpmc-ml}
\end{figure*}

In \Fig{zpmc-ml}, we show the cluster-wide $M^*/L_{i}$ as a function of $i$-band luminosity ($L_i$), halo mass within $R_{500}$ ($M_{500}$), and blue fraction.  Systematic vertical offsets and slope discrepancies in $M^*/L_i$ can be attributed to differences in the applied luminosity-to-mass conversion schemes (more on those below). Note also that the overall trends (or lack thereof) across the different formalisms (CMLR vs SED-fitting, varying IMF, varying SPS) are qualitatively consistent .  The values of $M^*/L_i$, $L_i$, and blue fraction correspond to the medians of their respective Monte Carlo distributions. The error ranges are taken to be between the 16th and 84th percentiles of these distributions.

\Table{ml-mm} summarizes the variation in cluster $M^*/L_i$ for each CMLR and SED-fit.
While the $M^*/L_i$ values span a broad range whose mean and width are contingent on the light-to-mass conversion used, there is little evidence that this variation varies with cluster luminosity (Pearson $R=-0.05$) or halo mass ($R=-0.09$).  When the data are organized against blue fraction, a weak, declining trend of $M^*/L_i$ with increasing blue fraction appears in both CMLR-based and SED-fitted outcomes ($R = -0.53$), as is expected, since ``red'' galaxies have more mass per unit red light.  

In their Figure 7, \citet{lea12} plot histograms of $M^*/L_{i+}$ for a sample of low-$z$ galaxy group members in the COSMOS survey, whose stellar masses were also derived from SED-fitting. The imaging (Cousins-$i+$ and Sloan $i$) is similar,
and the Chabrier IMF calibration is directly comparable to our MAGPHYS results. Therefore, we use their peak and variance of the $M^*/L_{i+}$ distribution to construct a ``COSMOS'' data point for comparison. That point is shown in the bottom row of \Fig{zpmc-ml}, where the published COSMOS data effectively extend our work into the regime of lower mass and bluer systems.

\Fig{zpmc-ml} (lower panels) shows the linear fits to the MAGPHYS-derived $M^*/L_{i}$ as a function of cluster mass (M$_{500}$) and blue fraction, including the COSMOS data point described above. These fits were performed using a bootstrapped (100 realizations) bisector method \citep{akritas96} which accounts for uncertainties in both observables, covariances between the observables, and intrinsic scatter. This method yields the following relations:

\be
\frac{M^*}{L_i} = -(2.10 \pm 2.01) ~ f_{blue} + (2.17 \pm 0.45),
\label{trendline-ml-vs-fblue}
\ee
and
\be
\frac{M^*}{L_i} = (-0.06 \pm 0.48) ~ \log_{10}\left(\frac{M_{500}}{10^{14}\,M_\sun}\right) + (1.71 \pm 0.23),
\label{trendline-ml-vs-m500}
\ee

with standard deviations of 0.20 and 0.21, respectively. While there is a weak trend toward decreasing $M^*/L_{i}$ with increasing blue fraction in \Fig{zpmc-ml}, that trend is not statistically significant (\Eq{trendline-ml-vs-fblue}). Further analysis of both rich clusters ($f_{blue} < 0.1$) and poor groups ($f_{blue} > 0.5$) could further constrain the relationship between the blue fraction and the global $M^*/L_{i}$, which ought to have a negative slope given that blue galaxies have lower $M^*/L_{i}$ than red galaxies.

\begin{table}[tb]
\caption{Variations in cluster M$_*$/L$_i$ for various models.}
\centering
\begin{tabular}{cc}
\hline\hline
Ref. & M$_*$/L$_i$\\
\hline
\citet{zib09} & $ 1.6 \pm 0.2$ \\
\citet{tay11} & $ 1.3\pm 0.1 $  \\ 
\citet{ip13} & $ 2.4 \pm 0.3 $   \\
MAGPHYS & $ 1.6 \pm 0.1 $ \\
\hline
\end{tabular}
\label{ml-mm}
\end{table}

\subsection{Variations due to L-M* Conversion Prescription}\label{sec:lmconvert}

As demonstrated in Figures \ref{zpmc-ml}, the derived $M^*/L_i$'s over the cluster mass range considered differ primarily between methodologies in zero-point and secondarily in scatter. The offsets range from Ta11 at the low end ($M^*/L_i \sim 1.3\pm0.1$), to Zi09 and MAGPHYS being comparable ($M^*/L_i \sim 1.6\pm0.2$), to Ip13 yielding the highest stellar mass-to-light ratios ($M^*/L_i \sim 2.4\pm0.3$). The sample variance of cluster $M^*/L_i$ is also widest for Zi09 and Ip13, but the differences are small. The cause for these zero-point shifts is a complex combination of underlying IMF choice, SPS modelling, SFH assumption, dust attenuation, etc. \citet{beh10} as well as Section 4.3 of \citet{lea12} address some of these effects. Of these, the choice of IMF generally preserves the power-law slope, but is otherwise a dominant source of systematic bias in $M_{\star}/L$. For instance, a Salpeter IMF typically gives stellar masses that are 0.25 dex above a Chabrier IMF, the Chabrier being, in turn, also 0.05 dex below a Kroupa IMF \citep{lea12}.

The IMF itself may also depend on environment. The Chabrier IMF was originally drawn from field (typically star-forming) galaxies, while the Salpeter IMF is thought to be a better representation of elliptical (more quiescent) systems. Hence, the most appropriate IMF choice might use a combination of a Chabrier IMF for blue systems and a Salpeter IMF for redder systems.   For reference, under our analytical framework, the \citet{bel03} CMLR with a diet-Salpeter IMF (\Table{cmlr-origins}) yields $M^*/L_i\sim2.7\pm0.2$ for our sample of clusters, which is in fact comparable to \citet{gon13}'s $M^*/L_I = 2.65$ (recall that our Chabrier-MAGPHYS result is $M^*/L_i\sim1.6\pm0.1$). 

The detailed physics of stellar evolution is also relevant as the SPS outcomes are especially sensitive to the treatment of bright thermally-pulsating asymptotic giant branch (TB-AGB) stars \citep[e.g.][]{beh10}. Note that Ta11 is based on the dated BC03 SPS package, which does not take TP-AGB into account. Zi09 and MAGPHYS use the CB07 SPS package, which does incorporate TP-AGB evolutionary tracks \citep{b07}.  Ip13 uses a more recent treatment of the TB-AGB phase, while also considering the effects of circumstellar dust. 

Part of the motivation for investigating an assortment of methods for computing total cluster stellar mass is to assess the absolute uncertainty, both statistical and systematic, in such an analysis.
To understand these complex systematics, a heurestic approach is to examine the variance in the predictions from a representative set of galaxy SED models varying in their underlying principles/assumptions. We take a first, by no means definitive, step towards realizing and characterizing the model-dependence of $M^*/L_i$ and $M^*/M_{500}$ (see \se{prevstudies} and \Table{cmlr-origins}). We demonstrate that, while absolute scalings and scatters are affected, overall trends exhibited by these stellar-mass quantities do not change appreciably.  Thus, we can conclude that, without prior information about the stellar populations (IMF, SFH, etc), a calculation of the total stellar mass from the mass-to-light conversion ($M^*/L_i = 1.3-2.7$) carries a $\sim$50\% uncertainty. If a specific IMF is well-motivated, \Table{ml-mm} demonstrates that this systematic uncertainty can be reduced to the $\sim$10\% level.
     
\subsection{Comparison with Previous Studies}\label{sec:prevstudies}

As reviewed in \se{intro}, baryon estimates in clusters often rely on heterogeneous light-to-mass transformations, such as applying a constant dynamical $M^*/L$ (e.g. \citealp{gon07,gon13, and10}) to the total cluster luminosities, or galaxy type- or luminosity-dependent $M^*/L$ scaling (e.g. \citealp{lin03,gio09,dai10,zha11,lag11, lin12}) in summing up the contribution of each galaxy to the cluster stellar mass despite observations that the fraction of blue to red galaxies in clusters can vary significantly \citep{geo11, lea12}. 

Following \citet{lea12}, we make no simplifying assumptions about cluster-wide mass-to-light ratios. Rather, each cluster galaxy and its five-band photometric information is modeled individually, applying either colour-dependent luminosity-mass conversion factors or directly deriving stellar masses from SED-fits. By testing a variety of light-to-mass conversion schemes, we can also quantify systematic effects on stellar mass derivations given different stellar population models, stellar formation/evolution scenarios, and galactic environments.  We also maximize photo-z information by constructing clusters with a probabilistic membership approach as described in \se{method}. Error margins on all measured parameters, conversion schemes, and membership uncertainties have been accounted in the overall error distribution of the final cluster stellar mass quotes.    
 
The first-order relation between cluster-wide $M^*/L_i$ and $M_{500}$, as captured in \Eq{trendline-ml-vs-m500} for the MAGPHYS results, indicates that the approximation of a constant mass-to-light ratio does {\it not} introduce systematic errors with respect to cluster size/mass. Furthermore, while one may expect a correlation between $M^*/L_i$ and the blue fraction, we find that the cluster-to-cluster scatter dwarfs this effect (see Eq.\ 8). The intrinsic spread of $M^*/L$ from the inhomogeneous nature of galaxy population cautions that cluster stellar mass estimates derived from constant light-to-mass factors should use  appropriate error margins to account for this population variance.  

The present work, \citet{lea12}, and \citealt{kra14} are among the few studies using a comparable methodology \footnote{\citet{kra14}'s sample overlaps with several of our clusters, for which they compute comparable total stellar content within $R_{500}$ when scaled to our mass-to-light ratios, as discussed in this section. Notably, \citet{kra14} use a more careful treatment of the BCG surface brightness profiles and caution that cModelMags underestimate total luminosity. See their Table 1.}. \citet{lea12} discern the stellar mass fraction of halos based on the COSMOS data using two methods:
(i) a cosmological simulation constrained by observations such as the stellar mass function, i.e. Halo Occupation Distribution (HOD)/abundance matching models, and direct X-ray group measurements, akin to this work. Both methods yield comparable results. \citet{lea12} infer galaxy stellar masses via individual SED-fits to multi-band COSMOS photometry assuming a Chabrier IMF, as we do with MAGPHYS; and (ii) \citet{geo11}'s reliable photo-$z$ based group membership catalogue to achieve high completeness above a stellar mass threshold. Two contrasting aspects of our approaches are:

\begin{itemize}
\item \citet{lea12} probe more distant systems (i.e. lowest redshift bin is $z\sim0.22 - 0.48$, compared to ours $0.04 < z < 0.1$);
\item \citet{lea12}'s sample of interest spans $M_{500}\sim10^{13}$ to $10^{14} M_\sun$ in the lowest redshift bin, though the theoretical HOD models extend the predicted $f_\star=M^*/M_{tot}$ to $M_{500}\sim10^{11} - 10^{15}M_\sun$.  Our sample spans the range: $M_{500}\sim10^{14} - 10^{15}M_\sun$.
\end{itemize}  

The stellar mass buildup between $z\sim0.3$ and $z\sim0.1$ is probably not substantial and thus our samples live in comparable regimes. 
The second point highlights the complementarity of these two studies, which span two orders of magnitude in mass when combined. Our study could also allow a direct comparison for the predicted $f_\star$ vs $M_{500}$ trends of HOD and abundance matching in high-mass regimes that we further address below. In practice, however, relatively shallow SDSS photometry may limit the accuracy of our $f_\star$ measurements (see below).  

\Fig{magphys-fstar-compare} compares the measured $f_\star$ vs $M_{500}$ relation from several recent studies with the results presented here, focusing exclusively on the MAGPHYS analysis. While varying the assumed IMF (i.e. use of various CMLRs) introduces systematic vertical offsets, it preserves the shape of the trend.

The MAGPHYS $M^*/M_{500}$ ratio exhibits a clear inverse correlation with halo mass $M_{500}$, corroborating numerous previous reports \citep[e.g.,][]{gio09,and10,zha11,lin12,lea12,gon13}. The best-fit power law is shown in red, with equation:
\be
M^*/M_{500} = 10^{3.49 \pm 1.05} (M_{500}/M_\sun)^{-0.38 \pm 0.08}. 
\label{magphys-ms-mtot-fit} 
\ee

We emphasize that our study cannot account for the entire stellar content of the cluster due to an important limiting factor arising from the photometric data themselves -- the SDSS cmodel\_mag luminosities are systematically underestimated, especially for bright galaxies with extended profiles (see \citealt{hall12}; \citealt{ber13}; \citealt{kra14}), translating into underestimated stellar masses. This factor would also affect the slope of \Eq{magphys-ms-mtot-fit} for smaller versus larger clusters; if the luminosities for brighter galaxies are more severely underestimated, and assuming their greater fractional influence in less massive clusters, then the true slope is likely steeper than that observed here. Nevertheless, the apparent consistency with literature (e.g., Figure \ref{magphys-fstar-compare}) for our crude $f_\star$ provides additional validation for our methods.  

We note that the dark blue points from \citet{gon13} include contributions from the `ICL' as defined in their particular study. The ICL-free data, which in the estimation of \cite{gon13} are $\sim25\%$ below the total cluster stellar mass, are plotted in cyan, constituting a more directly comparable measure to the rest of the works displayed.
\citet{gon13} is an extension of \citet{gon07}, with various improvements including a reduction of the (still constant) mass-to-light to $M^*/L_I=2.65$. \cite{lag11} lies nearly parallel to, but systematically above, our best-fit line, a discrepancy that can be partially ascribed to IMF-induced variations. The study by \citeauthor{lag11} uses the luminosity-determined mass-to-light ratio from \citet{kauf03}, which is based on a Kroupa IMF \citep{kro01}. However, Kroupa IMFs are known to yield only mildly larger stellar masses than Chabrier (0.05 dex), and much less than the Salpeter, whose expected coverage is also overplotted. 
The remaining discrepancy between our results and those of \cite{gon13} and \cite{lag11} may be due to incorrect subtraction of the background and improper modeling of the outer light profile, which is a known issue in SDSS ``cmodel'' magnitudes \citep{ber13,kra14} -- we will return to this point in \S5.4.
An additional source of the tension between our results and those of \cite{lag11} is our different estimates of the total halo mass. For clusters that overlap between our samples, we find factors of 10--30\% difference between our measurements of $M_{500}$, with measurements from \cite{lag11} being systematically lower (resulting in higher $f_\star$ estimates). Recent work has demonstrated that cluster masses based on \emph{XMM-Newton}-derived scaling relations may be biased low by a factor of $\sim$30\% \citep[see e.g.,][]{vonderlinden14}, which would explain most of the offset between our $f_{\star}$--$M_{500}$ relations.
Other factors, such as cluster sample variance and the use of oversimplified mass-to-light ratios and cluster member accounting, may also contribute to these systematic offsets. 

\begin{figure}[htb]
\center
\includegraphics[width=0.45\textwidth,trim=3.0cm 1.5cm 2.4cm 12.5cm]{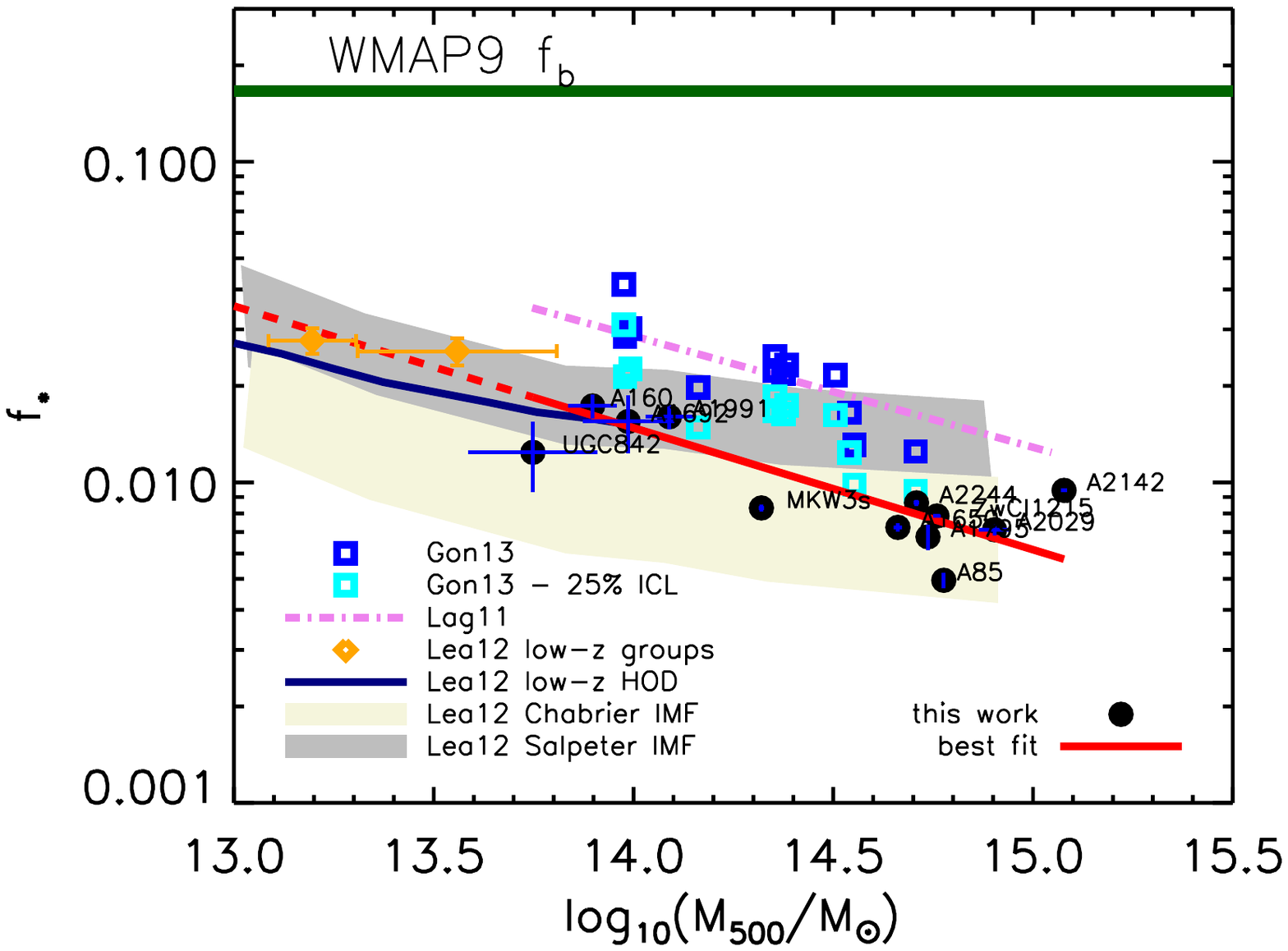}
\caption{Comparison of $f_\star \equiv M^*/M_{500}$ vs halo $M_{500}$ from previous studies with our data (MAGPHYS) shown in black. The reference abbreviations are: Gon13 = \citealt{gon13}, Lag11 = \citealt{lag11}, Lea12 = \citealt{lea12}. For clarity, some error bars, such as those on the Gon13 points, have been omitted. Due to limitation in SDSS photometry, our results likely underestimate the absolute stellar fraction. The cyan squares of Gon13 have a 25\% ICL contribution reduction but remain more star-rich than our stellar fractions based on the Chabrier IMF. Lag11's displacement above our work is partly due to systematic offsets in stellar mass (IMF) and total mass between the two works (see discussion in \S5.3). Orange diamonds are the binned results from Lea12's low-z ($z = [0.22, 0.48]$) groups, which are deemed most comparable to our own study. The HOD fit to the same redshift range is plotted in navy blue. Abundance matching techniques assuming Chabrier and Salpeter IMFs are extensible over a large dynamic range of halo masses, and yield the respective error margins in beige and gray (see figure 5 in \citealt{lea12}). The best-fit line to our data (in red) has equation $f_{star} = 10^{3.49 \pm 1.05} (M/M_{500})^{-0.38 \pm 0.08}$ which, when extrapolated to low halo masses (dotted red), is entirely consistent with \citet{lea12}'s observations for groups (orange points). Also note the consistency between our high-mass cluster results and those predicted from fitting cosmological simulations, corroborating the hypothesis that $f_\star$ declines as a single power law from low to high-mass halos. See text for further interpretation.}
\label{magphys-fstar-compare}
\end{figure}

\Fig{magphys-fstar-compare} shows that our results naturally extend the power-law of \citeauthor{lea12}'s lower-mass galaxy groups over the range $M_{500} \sim 10^{13} - 10^{15} M_\sun$, essentially consistent with the Charbrier extrapolation of models from abundance matching in \citeauthor{lea12}. This suggests that 1) the relevant physics of galaxy groups transitions smoothly to massive clusters, hence both may be studied under the same framework; 2) if universal baryon fractions, $f_b = f_{gas} + f_\star$, are constant with cluster scale to first order, then the trend of declining stellar mass fraction in increasingly more massive halos implies that $M^*/M_{gas}$ is lower for higher-mass cluster systems. In turn, this suggests that baryon cooling would be less efficient in deeper potential wells; and 3) the lower absolute scaling of total stellar fractions than previous expectations (due primarily to different choices of IMF) makes explaining the baryon fraction deficiency with respect to WMAP's cosmic measurements increasingly challenging. These interpretations should be read with the caveat that our absolute stellar masses likely represent underestimates of their true values.

\subsection{Methodological Uncertainties \\
and Incompleteness}\label{sec:uncertain}

Stellar mass estimates for individual galaxies remain poorly constrained \citep{courteau14}.
For galaxy clusters, systematic and statistical pitfalls due to stellar evolution modeling \citep{beh10} combined with membership uncertainty make the assessment of the total stellar mass even less certain.
Still, once it is recognized that systematics dominate over statistical uncertainties in stellar mass measurements (see section 4.3 in \citet{lea12}), the latter can be assessed given the former as background and the error margins can be separately decoupled. For instance, errors attributable to stellar population modeling systematics can be probed by repeating our methodology for a number of different stellar light-to-mass conversion techniques, bootstrapping over all statistical errors associated with the particular technique. The range of results for a representative pool of such techniques indicates the nature of the systematic variation due to the assumptions of the conversion scheme. In this work, we have mostly explored these systematic effects by supplementing MAGPHYS calculations with several popular CMLRs.

\begin{figure}[htb]
\center
\vspace{-160pt}
\includegraphics[width=0.46\textwidth,trim=2.2cm 0cm 2.2cm 0cm]{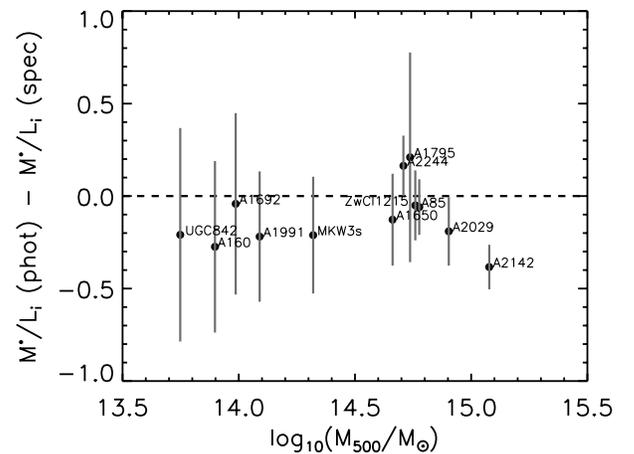}
\vspace{-30pt}
\caption{Expected stellar mass-to-light ratio bias if the $z_{phot}$ probability distributions used in this work were not supplemented by $z_{spec}$ measurements. For this exercise, we re-construct each cluster using only the subset of candidate galaxies with both spectroscopic and photometric redshifts. We sum MAGPHYS-derived stellar masses via the probabilistic methodology described in \se{method}, once using exclusively spectroscopic redshifts, and once using photometric. For the subsample with definitive membership assignment using spectroscopic redshift measurements, photometric redshift alone appears to yield systematically lower mass-to-light ratios, by $\sim0.15 M_{\odot}/L_{\odot}$, which is similar to the uncertainty in the calculated $M^*/L_i$.}
\label{ml_s-p}
\end{figure}

The uncertainty due to the cluster member accounting (see \se{method}) is now examined.  Deprojection effects amount to less than a 15\% downward adjustment from selecting galaxies in a cylinder to a sphere in redshift space \citep{lea12}. 
In \S3 we further quantify this bias as a function of spectroscopic completeness, showing that our cluster membership algorithm is likely biased by $<$20\% for systems with spectroscopic completeness $>$10\%.
The photometric redshift probability distribution has an uncertainty that can also be characterized by applying different prob($z$) catalogues (if available to the same field) and inspecting the resulting variance. This error source is limited by the reliability of the externally-supplied form of prob($z$) for a cluster field. 
A first-order estimate of the error associated with using photometric redshifts can be made by isolating the spectroscopic subset of cluster candidates and performing the entire analysis on these galaxies alone. These galaxies have unambiguous true redshifts and membership designations, making their aggregation a benchmark for comparing the accuracy of calculations based on photometric redshifts. We then re-compute the parameters using our statistical treatment of $z_{phot}$ estimates (\se{method}) and compare the two outcomes. \Fig{ml_s-p} shows the differences between the photometric and spectroscopic redshift-based calculations of median  $M^*/L_i$ as function of halo mass, $M_{500}$.  The mass-to-light ratios inferred using only photometric redshifts tend to be underestimated with respect to those computed using spectroscopic redshifts. The median bias of $\sim0.15 M_{\sun}/L_{i,\sun}$ is similar in magnitude to the typical uncertainty for a given cluster $M^*/L_i$.

Another possible source of bias, as we have alluded to in previous sections, is the use of ``cmodel'' magnitudes from the SDSS pipeline. \cite{ber13} showed that roughly 20\% of the total stellar mass density at $z\sim0.1$ is missing in SDSS cmodel magnitudes, with the differences being largest at the high mass end. Since galaxy clusters tend to have a higher fraction of massive galaxies, this bias could be even higher for our sample. In the most extreme case -- that of a galaxy cluster with $M_{500} \sim 10^{14}$ M$_{\odot}$, where the BCG contributes a substantial fraction of the total mass -- this could bias our estimate of M$_*$ low by 35--40\%. In an attempt to determine the maximal effect of this bias on the measured M$_*$/L$_i$, we can assume that the stellar populations in the outer halo of galaxies (where ``cmodel'' magnitudes underestimate the flux) are exceptionally young ($M_*$/L$_i = 1.0$). This extreme scenario, combined with the maximal bias of 35--40\% mentioned above, would lead to a bias in M$_*$/L$_i$ of 15\%. Of course, this scenario is also unrealistic, as \cite{roediger11} has shown that Virgo cluster galaxies have relatively flat age and metallicity gradients in their outskirts. However, this calculation shows that any bias due to the use of ``cmodel'' magnitudes rather than those, for example, presented in \cite{ber13}, is likely $<$15\%.

Mass incompleteness, due to the magnitude-limited nature of surveys, is also a source of concern. Luminosity function fits \citep[e.g.,][]{schechter76} have often been invoked to estimate the contribution of undetected galaxy members below a luminosity threshold . We choose not to extrapolate our LFs because 1) faint-end slopes are poorly constrained, and 2) the gain in accuracy for our present purpose is minimal. To substantiate the latter claim, note that, being nearby ($z<0.1$), our clusters are well sampled by the SDSS: our magnitude limit of $r < 22$ mag corresponds to 0.01L$_{\star}$.

\begin{figure}[htb]
\center
\includegraphics[width=0.49\textwidth,trim=2.3cm 2cm 1.5cm 12cm]{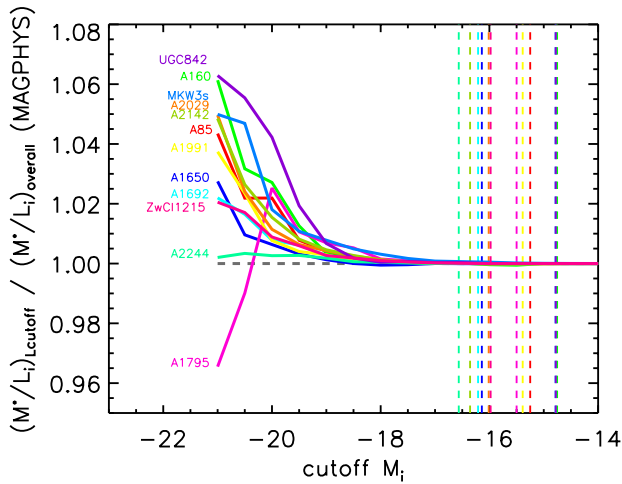}
\caption{Convergence of mass-to-light ratios to their final values as a function of absolute magnitude limit of cluster galaxy detection. Having all cluster galaxies brighter than $M_i = -19$ already anchors the overall $M^*/L_i$ to within 1\%. The convergence is settled by $M_i = -17$, which is not a stringent requirement for nearby systems (see text). Vertical dashed lines mark the approximate positions of limiting absolute $i$-magnitudes corresponding $r<22$ for each cluster redshift, assuming $r-i\approx0.3$ }
\label{magphys-ml-converge}
\end{figure}

Recall that our primary goal is to characterize cluster-wide stellar mass-to-light ratios which, being a weighted average, does not require a complete accounting of all galaxies present (as long as the missing galaxies do not deviate significantly from the mean). In order to determine how deep a given survey must be for the uniform-$M^*/L_i$ approximation to be suitable, we show in \Fig{magphys-ml-converge} the derived $M^*/L_i$ as a function of the survey depth. Firstly, note the range spanned by the ordinate. Detection of all galaxy members brighter than $M_i=-19$  ensures that $M^*/L_i$ falls within 1\% of the final value, despite the fact that the vast majority of the galaxies we detect in each cluster are fainter than this limit. For a cluster at $z=0.1$, this corresponds to a magnitude limit of $m_i = 19.3$ mag, which is certainly attainable by SDSS. The rapid convergence of this plot, which occurs well clear of our survey limits, suggests that faint-end incompleteness does not affect our conclusions regarding cluster $M^*/L$.

We conclude our discussion on uncertainties by advocating the use of multi-wavelength photometry (such as COSMOS) for stellar mass estimates whenever possible. In the absence of exceptionally well-sampled photometry spanning the optical through IR \citep[e.g., COSMOS, CLASH;][]{lea12,postman12}, we recommend at least three photometric bands which span the 4000\AA\ break, which will allow a coarse modeling of the SED, yielding improved constraints on $M^*/L_i$ over CMLRs.

\section{Conclusion}\label{sec:conclusions}

We have presented an investigation of the stellar mass budget for galaxy clusters in the total mass range $M_{500} \sim 10^{13} - 10^{15} M_\sun$.  We specifically addressed the distribution of overall cluster mass-to-light ratio values, and the total stellar mass in galaxies as a fraction of total halo mass for clusters of various sizes. Towards this end, we have developed a bootstrapping algorithm for cluster stellar mass accounting on a galaxy-by-galaxy basis in the absence of complete spectroscopic field coverage as an alternative to statistical, flux-based background subtraction methods. Our method can be applied to any photometric redshift catalogue and stellar light-to-mass conversion models, making it valuable in assessing model-dependent systematic uncertainties. We have repeated our tests with several widely-cited colour-mass-to-light relations (CMLRs) and also using full SED-fitting with the program MAGPHYS utilizing five-band optical photometry from SDSS. Our results are summarized below:

\vskip 0.1 in
\noindent{}1) We find no strong evidence for first-order dependence of cluster-wide $M^*/L_i$ on halo mass. Cluster mass-to-light ratios span relatively narrow ranges ($\pm$10\%) with the absolute level and intrinsic variance set by the specifics of the stellar population models (see \Fig{zpmc-ml} and \Table{ml-mm} for these results). Not surprisingly, $M^*/L_i$ is found to weakly correlate (Pearson $R=-0.53$) with cluster blue fraction, though an expanded cluster sample is required for a more precise quantification of this statement. 

\vskip 0.05 in
\noindent{}We have shown that the acceptable range of $M^*/L_i$ ratios varies as a function of the preferred IMF and SPS packages, but those are essentially independent of the halo mass. For the popular Chabrier IMF coupled with the CB07 SPS prescription, we advocate a MAGPHYS value of $M^*/L_i \approx 1.7 \pm 0.2$ for galaxy clusters.  For a (diet) Salpeter IMF, a higher $M^*/L_i$ value of $\sim2.7$ may be more appropriate for the same clusters. The truth likely lies in between.    

\vskip 0.1 in
\noindent{}2) Despite limitations of our present total stellar mass estimates, which make them imperfect for accurate cluster stellar content accounting, we measure a strong correlation between the cluster stellar-halo mass fraction ($f_{\star}$) and halo mass. Such a trend supports the emerging consensus that star formation is less efficient in deeper potential wells (e.g. \citealt{lin03,gon13}) . Our SED-fitting analysis based on MAGPHYS also yields: $f_\star = 10^{3.49 \pm 1.05} (M/M_{500})^{-0.38 \pm 0.08}$ in agreement with previous studies. The zero-point of this relation depends on the adopted CMLR or SED-fit for the stellar mass determinations, but the slope is preserved. A declining stellar mass fraction with increasing halo mass has yet to be reconciled with theoretical predictions. See \citet{gon13} and discussion therein. 

\vskip 0.1 in
Our study is similar in spirit to that of \cite{lea12}, who compute stellar masses and mass-to-light ratios for galaxy groups in the COSMOS survey. As illustrated in \Fig{magphys-fstar-compare}, our results form a smooth extension (to higher mass) of their work, corroborating the notion that stars may contribute even less to the cluster baryon budgets than previously expected, especially in massive halos.

We have restricted our study to nearby galaxy clusters, but the methodology described here can be easily extended to a wide range of redshifts and data sets, provided that the photometric redshifts are reliable and the photometry is relatively deep ($M_{i,lim} < -19$ mag).

\section{Acknowledgement}

We are grateful to the anonymous referee for their thoughtful suggestions, which improved the content and clarity of this paper. Y.~S. and S.~C. are grateful to the Natural Science and Engineering Research Council of Canada for funding via an Undergraduate Summer Research Award and Research Discovery Grant, respectively, which made this study possible. M.~M. acknowledges support provided by NASA through a Hubble Fellowship grant from STScI.  Comments from Anthony Gonzalez on an earlier draft were most helpful. We also thank Larry Widrow, Joel Roediger, Carlos Cunha, and John Moustakas for constructive suggestions at various stages of this project. 

Funding for SDSS-III has been provided by the Alfred P. Sloan Foundation, the Participating Institutions, the National Science Foundation, and the U.S. Department of Energy Office of Science. The SDSS-III web site is http://www.sdss3.org/.

SDSS-III is managed by the Astrophysical Research Consortium for the Participating Institutions of the SDSS-III Collaboration including the University of Arizona, the Brazilian Participation Group, Brookhaven National Laboratory, Carnegie Mellon University, University of Florida, the French Participation Group, the German Participation Group, Harvard University, the Instituto de Astrofisica de Canarias, the Michigan State/Notre Dame/JINA Participation Group, Johns Hopkins University, Lawrence Berkeley National Laboratory, Max Planck Institute for Astrophysics, Max Planck Institute for Extraterrestrial Physics, New Mexico State University, New York University, Ohio State University, Pennsylvania State University, University of Portsmouth, Princeton University, the Spanish Participation Group, University of Tokyo, University of Utah, Vanderbilt University, University of Virginia, University of Washington, and Yale University.


\vspace{+10pt}

\end{document}